\AtBeginDocument{%
  \providecommand\BibTeX{{%
    \normalfont B\kern-0.5em{\scshape i\kern-0.25em b}\kern-0.8em\TeX}}}
\documentclass[sigconf]{acmart}
\acmConference[ESEC/FSE 2020]{The 28th ACM Joint European Software Engineering Conference and Symposium on the Foundations of Software Engineering}{8 - 13 November, 2020}{Sacramento, California, United States}
\usepackage[utf8]{inputenc}
\usepackage{dblfloatfix}  
\usepackage{multicol}
\usepackage{graphicx}
\usepackage{wrapfig}
\usepackage{colortbl}
\usepackage[ruled,vlined]{algorithm2e}
\usepackage[final]{listings}
\usepackage[framemethod=tikz]{mdframed}
\usepackage{tcolorbox}
\newtcolorbox{blockquote}{colback=red!5!white,boxrule=0.4pt,colframe=red!50!black,fonttitle=\bfseries}

\title{How to Improve AI Tools (by Adding in SE Knowledge): \\Experiments with the TimeLIME  Defect Reduction  Tool}
\author{Kewen Peng}
\affiliation{%
  \institution{North Carolina State University}
  \city{Raleigh}
  \country{USA}}
\email{kpeng@ncsu.edu}

\author{Tim Menzies}
\affiliation{%
  \institution{North Carolina State University}
  \city{Raleigh}
  \country{USA}}
\email{tjmenzie@ncsu.edu}

\date{January 2020}

\begin{document}
\begin{abstract}
AI algorithms are being used with increased frequency in SE research and practice. Such algorithms are usually commissioned and certified
using   data
from outside  the SE domain.  Can we assume that such algorithms can be used   ``off-the-shelf'' (i.e. with no modifications)? To say that another way,
are there special features of SE problems that suggest a different and better way to use  AI tools?

To answer these questions, this paper reports 
experiments with TimeLIME, a variant of the   LIME explanation algorithm from KDD'16.  LIME can offer recommendations on how to change
 static code attributes in order to reduce the number of defects in
the next software release.   That version of LIME used
an internal weighting tool to decide what attributes to include/exclude in 
those recommendations.
TimeLIME improves on that weighting scheme using  the following  SE knowledge:
 software comes in releases; an implausible change to software is something that has  never been  changed in prior releases; so it is better to use
plausible changes, i.e. changes with some precedent in the prior
releases.
By   restricting
recommendations to just the frequently changed attributes,   TimeLIME   can produce (a)~dramatically better explanations of what causes defects and (b)~much better recommendations on how to fix
buggy code. 

Apart from these specific results about defect reduction and TimeLIME, the more general point of this paper
is that our community should be  more careful about using off-the-shelf AI tools, without first applying SE knowledge.
As shown here, it may not be a complex matter to apply that knowledge.
Further,
once that SE knowledge is applied, this can result in dramatically better systems.

\end{abstract}
\maketitle
\section{Introduction\label{intro}}

This paper finds and fixes a flaw in a widely cited AI explanation generation method, LIME (first presented at KDD'16). In theory, LIME can be used to find code changes that 
make software less buggy in the next release. 
In practice,  when we tried doing that, we found that the classic LIME model was generating
 surprising and unprecedented recommendations. 
Specifically, classic LIME kept   suggesting changes that  had
 never been seen before in the history of the project. 

% make implausible recommendations that have not been seen in prior releases of the code.

When we first observed this, our initial response was quite favorable. Perhaps, we thought,
LIME would offer novel and powerful suggestions  that would lead to greater defect reductions
than ever seen before.
However, as shown in this paper, classic LIME's recommendations are sub-optimal. This paper presents TimeLIME which is  a  version of LIME that restricts its explanations to
the attributes that change the most. On experimentation, TimeLIME's explanations were seen to be:
\begin{itemize}
\item {\em Smaller }: TimeLIME restricts itself to the $M=5$
most changed attributes.  Classic LIME, on the other hand, uses dozens more attributes.
\item {\em Easier to apply:} The fewer the recommendations, the quicker it is to act on those recommendations. 
\item {\em Better explanations}: 
The recommendations from  TimeLIME are associated with a much larger reduction
in defects than classic LIME.
\end{itemize}
While TimeLIME is certainly a useful tool for proposing code changes, we argue that 
this is less important than {\em how this result was generated}.
AI algorithms are being used with increased frequency in SE research and
in SE industrial practice. If  these AI tools
are used ``off-the-shelf'' (i.e. with no modifications), then that  assumes that
the  problems used to commission and certify these AI
algorithms  are relevant to SE problems. 
The results of this paper suggest that such assumption can be very dubious.
As shown below, the performance of standard AI tools can be enhanced dramatically just by applying
some SE knowledge. Specifically, the contribution of this paper is to improve classic LIME via three items of SE knowledge:
\begin{itemize}
\item
Software comes in releases. 
\item
An implausible change to software is something that has never been changed in prior releases. 
\item
It is  better to use plausible changes, i.e. changes with some precedence in the prior releases.
\end{itemize}

\begin{table*}[t!]
\scriptsize
\begin{tabular}{|l|l|l|}
\hline
\rowcolor[HTML]{C0C0C0} 
Metric  & Name                        & Description                                                                                                     \\ \hline
amc     & average method complexity   & Number of JAVA byte codes                                                                                       \\ \hline
avg\_cc & average McCabe Average      & McCabe’s cyclomatic complexity seen in class                                                                    \\ \hline
ca      & afferent couplings          & How many other classes use the specific class.                                                                  \\ \hline
cam &
  cohesion amongst classes &
  \begin{tabular}[c]{@{}l@{}}Summation of number of different types of method parameters in every method divided by a multiplication \\ of number of different method parameter types in whole class and number of methods.\end{tabular} \\ \hline
cbm     & coupling between methods    & Total number of new/redefined methods to which all the inherited methods are coupled                            \\ \hline
cbo     & coupling between objects    & Increased when the methods of one class access services of another.                                             \\ \hline
ce      & efferent couplings          & How many other classes is used by the specific class.                                                           \\ \hline
dam     & data access                 & Ratio of private (protected) attributes to total attributes                                                     \\ \hline
dit     & depth of inheritance tree   & It’s defined as the maximum length from the node to the root of the tree                                        \\ \hline
ic      & inheritance coupling        & Number of parent classes to which a given class is coupled (includes counts of methods and variables inherited) \\ \hline
lcom    & lack of cohesion in methods & Number of pairs of methods that do not share a reference to an instance variable.                               \\ \hline
locm3 &
  another lack of cohesion measure &
  \begin{tabular}[c]{@{}l@{}}If $m$, $a$ are the number of methods, attributes in a class number and $\mu(a)$ is the number \\ of methods accessing an attribute, then lcom3 = $((\frac{1}{a}\sum_{j}^{a} {\mu(a_j)})-m)/(1-m)$ \end{tabular} \\ \hline
loc     & lines of code               & Total lines of code in this file or package.                                                                    \\ \hline
max\_cc & Maximum McCabe              & Maximum McCabe’s cyclomatic complexity seen in class                                                            \\ \hline
mfa     & functional abstraction      & Number of methods inherited by a class plus number of methods accessible by member methods of the class         \\ \hline
moa     & aggregation                 & Count of the number of data declarations (class fields) whose types are user defined classes                    \\ \hline
noc     & number of children          & Number of direct descendants (subclasses) for each class                                                        \\ \hline
npm     & number of public methods    & Npm metric simply counts all the methods in a class that are declared as public.                                \\ \hline
rfc     & response for a class        & Number of methods invoked in response to a message to the object.                                               \\ \hline
wmc &
  weighted methods per class &
  A class with more member functions than its peers is considered to be more complex and therefore more error prone \\ \hline
defect  & defect                      & Boolean: where defects found in post-release bug-tracking systems.                                              \\ \hline
\end{tabular}

\vspace{5mm}
\caption{The C-K OO metrics used in defect prediction. The last variable "defect" is the dependent variable.}
\label{ck}
\end{table*}
Based on the experience of this paper, we caution that  our community should be more careful about using off-the-shelf AI tools, without first tuning them with SE knowledge.
As shown here, it is may not be a complex matter to apply that knowledge. Further,
once that SE knowledge is applied, this can result in dramatically better systems.

\noindent This paper is structured
around the following research
questions.

\noindent\textbf{RQ1: Are all explanations precedented?}
    \begin{blockquote}
    \noindent
    \textbf{Answer 1}: Widely-used explanation algorithms (classic LIME) do not restrict themselves to explanations with precedence in the historical record of a project.
    \end{blockquote}
We view this first result as a potential flaw in classic-LIME.
As shown by \textbf{RQ3}, better
explanations can be found using precedented explanations.

\textbf{RQ2: Do developers prefer   precedented explanations?}
    \begin{blockquote}
    \noindent
    \textbf{Answer 2}: Of all the planners studies here, developers are less likely to perform the plans proposed by classical LIME than TimeLIME. That is to say, the precedented explanations are more favored by developers.   
    \end{blockquote}
\textbf{RQ3: Are precedented explanations better at defect reduction?}
    \begin{blockquote}
    \noindent
    \textbf{Answer 3}: 
    TimeLIME's precedented explanations are associated with  greater defect reduction.
    They are also easier for developers to apply.
    \end{blockquote}

The rest of this paper is structured as follows.
\S\ref{background} discusses  defect prediction and trends in the explanation literature.  \S\ref{K-test} shows our method for ranking different planning
methods.   \S\ref{experiment}     
describes  experiment and  the  datasets, predictive model, and planners used in this work. \S\ref{result} reports our result.  The credibility and reliability of our conclusions is discussed by \S\ref{threat}. Finally, we offer conclusions and discuss future work in \S\ref{futurework}and \S\ref{conclusion}.

\subsection{Data Availability}
All the data and scripts used in this paper are freely available online at \url{http://github.com/anonymous12138/FSE2020}.
\section{Background and Related Work}\label{background}
\subsection{Defect Prediction}

The case study of this paper comes from defect prediction and planning. This kind of analysis is discussed
in this section.

During software development, the testing process  often has some resource limitations.
For example, the effort associated with coordinated human effort across a large code
base can   grow exponentially with the scale of the project \cite{fu2016tuning}.

\begin{figure*}[!b]
~\hrule~
\begin{minipage}{.49\linewidth}
\includegraphics[height = 4.5cm,width=.9\linewidth]{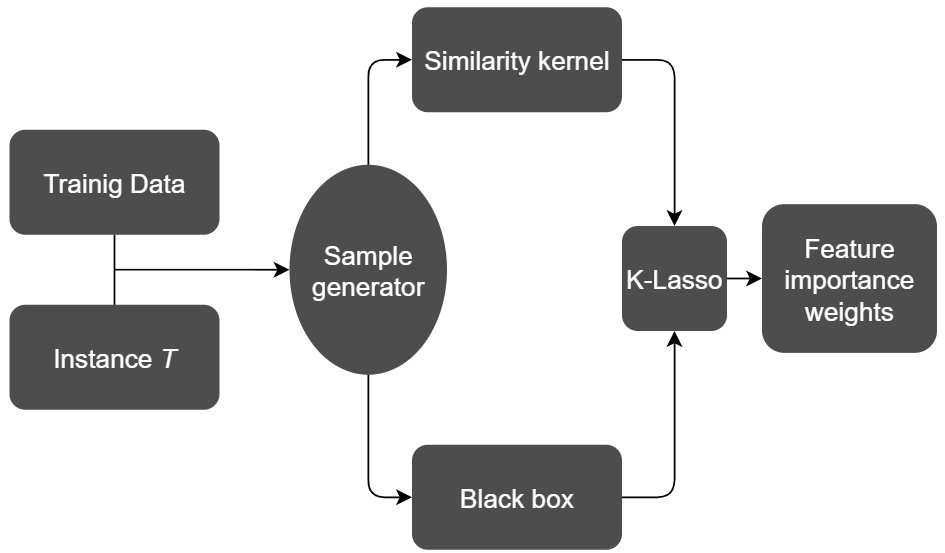}
\end{minipage}
\begin{minipage}{.49\linewidth}
\small
\begin{itemize}
\item
LIME is designed to be an add-on to other AI systems
(e.g., neural network, support vector machine, and so on).
Hence, it treats those AI tools as a 
``\emph{black box}'' that is queried within its processing.
\item
Within LIME, some  \emph{sample generator} is used to generate synthetic data which later gets passed to  the  \emph{black box}  and a \emph{similarity kernel}, along with the original \emph{training data}. 
\item
The \emph{similarity kernel} is an instrument used to weight the prediction results of \emph{training data} returned by the \emph{black box} by how similar they are to the \emph{instance T}. 
\item
The \emph{K-Lasso} is the procedure that learns the importance weights from the $K$ features selected with Lasso using a class of linear models.
\end{itemize}
\end{minipage}
~\hrule~
\caption{Inside LIME. From~\cite{ribeiro2016should}. The feature importance weights are passed to Algorithm \ref{simpleflip} and \ref{refineflip}, as later elaborated in \S\ref{planner}.
For a sample of the output feature importance
weights, see Figure~\ref{fig:lime}.}\label{inside}
\end{figure*}

Hence, to effectively manage resources, it is common to match the quality assurance (QA) effort
to the perceived criticality and bugginess of the code. 
Since every decision is associated with a human and resource cost to the developer team, it is impractical and inefficient to distribute equal effort to every component in a software system\cite{briand1993developing}.   Learning 
defect prediction (using data miners)
from static code attributes (like those shown in Table \ref{ck}) is one very cheap way to ``peek'' at the code and decide
where to spend more QA effort.

Recent results show that software defect predictors  are also competitive widely-used  automatic methods.  
 Rahman et al. ~\cite{rahman2014comparing} compared (a) static code analysis tools FindBugs, Jlint, and PMD with (b) defect predictors (which they called ``statistical defect prediction'') built using logistic  regression.
No significant differences in   cost-effectiveness were observed.
Given this equivalence, it is significant to
note that  defect prediction can be quickly adapted to new languages by building lightweight parsers to extract  code metrics. The same is not true for static code analyzers - these need extensive modification before they can be used in new languages.
Because of this ease of use,  and its applicability to many programming languages, defect prediction has been   extended  many ways including:
\begin{enumerate}
\item Application of defect prediction methods to locating code with security vulnerabilities~\cite{Shin2013}.
    \item Understanding the factors that lead to a greater likelihood of defects such as defect prone software components using code metrics (e.g., ratio comment to code, cyclomatic complexity) \cite{menzies10dp, menzies07dp} or process metrics (e.g., recent activity).
    \item Predicting the location of defects so that appropriate resources may be allocated (e.g.,~\cite{bird09reliabity})
    \item Using predictors to proactively fix  defects~\cite{arcuri2011practical}
    \item Studying defect prediction not only just release-level \cite{chen2018applications} but also change-level or just-in-time \cite{commitguru}.  
    \item Exploring ``transfer learning'' where predictors from one project are applied to another~\cite{krishna2018bellwethers,nam18tse}.
    \item Assessing different learning methods for building predictors~\cite{ghotra15}. This has led to the development of hyper-parameter optimization and better data harvesting tools \cite{agrawal2018wrong, agrawal2018better}. 
\end{enumerate}

This paper extends defect prediction and planning in yet another way: exploring the  trade-offs between explanation and planning and the performance of defect prediction models. But beyond the specific scope of this paper, there is nothing in theory stopping the application of this paper to all of the seven areas listed above (and this would be a fruitful area for future research).

\subsection{Planning as Explanation Generation}
% why lack
In principle, once it is known how a conclusion is reached, we can query that method to
find out how to change something in order to reach better conclusions.
This intuition is the core of explanation-based planners. Such planning can proceed as follows: 
\begin{enumerate}
\item
Use standard means
to build models that make predictions;
\item
Perform what-if queries across those
models to find plans on how to change the prediction. 
\end{enumerate}
Depending on the nature of the model,
those what-if queries can be very slow (e.g., Monte Carlo simulations over a neural net model)
or very fast (e.g., just use the  attributes with the largest  $\beta$ coefficients found by linear regression). The LIME method described later in this paper is an example of a very fast method.

\subsection{Explaining ``Explanations''}

In our experience,
software developers prefer a transparent decision-making model in which 
some valid rationale is provided behind each decision so that they may argue the merits of such decisions. From the perspective of transparency, the term "explanation" or "interpretability"   refers to the extent of the human comprehension of a given AI system or the decisions made by it.

As documented in their 2019 literature review, Mueller et.al.~\cite{mueller2019explanation} observes that research on formal and computational models of explanation is truly vast
and dates back many centuries.
Formal explorations of the concept of explanation can be found  in the ``fourth-figure'' of Aristotle~\cite{leake91}. Written in the 19th century,
explanation was characterized by Charles Sanders Peirce  as follows: "The surprising event C is observed. But if A were true, C would be a matter
of course. Here, there is reason to suspect that A is true"~\cite{AAAI_1990}.
Mueller et al. acknowledge Peirce's historical leadership in this field but warn
that Peirce's formalism misses at least two 
important features: 
specifically, problem formulation and problem resolution. They comment
that mapping an explanation back to action is  ``is where the hard work of explanation occurs, and that the (Peirce) model is not specific about what is involved in these steps.''

In the 1980s and 1990s,  a further  nuance  was introduced in the
the concept of explanation.  Researchers exploring knowledge-based systems found that it
was not enough to view explanations as  a pretty print of a trace of
some inference procedure. Even when the inference trace was across very succinct domain-specific languages, researchers like Leake and Clancey
were surprised to see that different users wanted different kinds of explanations~\cite{leake91,clancey94}. They concluded that 
explanation is a separate problem-solving task to inference. In their view, explanation is a procedure that customizes what to be reported according to the task at hand.  Explanations, in this modern view, is context-specific: and "the context of the current situation can
significantly affect the purpose and therefore the content of an explanation"~\cite{menzies02}.

Explanation research stresses the need for some form of 
{\em plausibility}
operator in order to prevent the presentation of bogus explanations~\cite{menzies02}. Consider two explanations
for "the grass is wet". This might  have happened because either because (a)~it rained last night or  
(b)~the lawn sprinkler has been left on.  Hence, explaining  "wet grass" using "rain" is a possible,
but not necessarily a certain, inference. Plausibility operators~\cite{menzies02}.
can be used to assess and cull weaker explanations. Returning to the grass example, if this was a lawn in
Albuquerque (which is a desert city) and if the time was high summer (which is usually very dry) then a plausibility operator would favor explanations that use ``sprinkler'' over ``rain'' since the latter is unlikely
in a desert city in summertime.

\begin{figure}[]
\tiny
\includegraphics[width=1.0\linewidth,height=4cm]{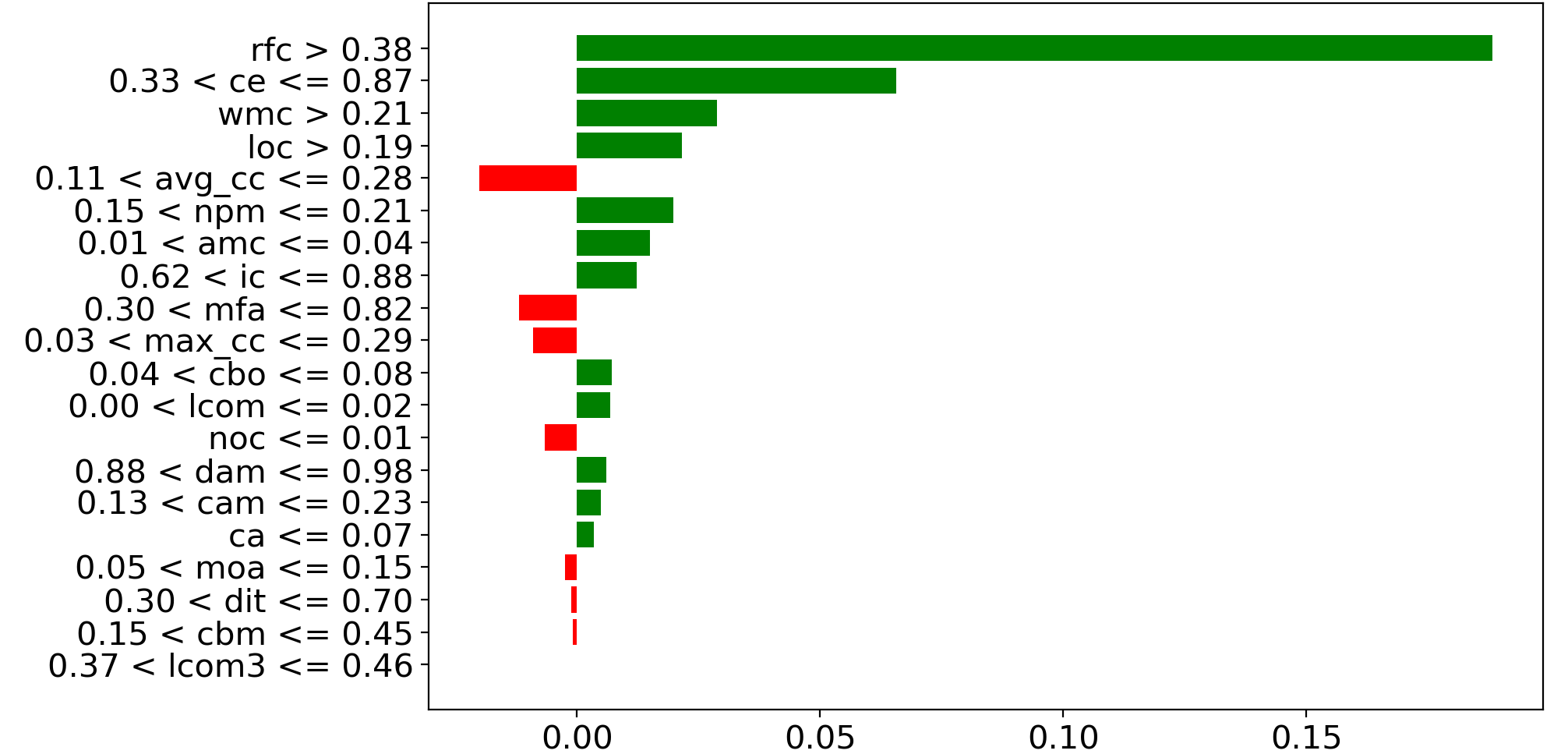}
\caption{
An example of
output generated by 
Figure~\ref{inside} when applied
to the data sets of the form
of Table~\ref{ck}.
The y-axis shows the feature name and the confidence interval during which the explanation stays effective. The x-axis indicates the importance weight of each attribute. The prediction label of this instance is 1 (defective), and the weights show how each feature contributes to the prediction. A positive weight means the feature encourages the classifier to predict the instance as a positive label (1), and vice versa. Larger weights indicate greater feature importance in terms of the prediction value based on that feature weighted by a similarity kernel.}
\label{fig:lime}
\end{figure}

It is insightful to review the LIME explanation algorithm in the context of the above  paragraphs:
\begin{itemize}
\item
In terms of mapping explanation to action, LIME takes the view that a ``good''
explanation is one that can change the class of some test instance. To that end, 
LIME builds a linear approximation model from examples near the test instance. Using that model, LIME learns what needs to be changed within the test instance in order to change the class variable of that instance (see Figure~\ref{inside}). The output of LIME is hence a set of attribute ranges, sorted
by how much those ranges could alter a class label (see Figure~\ref{fig:lime}).
\item
In keeping with the work from the 80s and 90s, LIME is a context-specific 
explanation system. Unlike data miners (that generate one model to be applied to all test instances), LIME generates a different explanation for each specific test instance. 
\item
As to the plausibility operator, LIME uses its  its own internal weighting  scheme to rank explanations. The argument of this paper is that, when generating explanations over
multiple consecutive releases of some software system,  a useful plausibility operator is to restrict explanations to those changes seen in  recent historical record of the project.
\end{itemize}

% The result returned by LIME is the importance weights of all $K$ features.  In summary, the algorithm
% builds a local linear regression model use examples ``similar'' to some test instance.
% Here, ``similar'' is defined using the {\em similarity kernel}. Some researchers apply uniform weights to all instances in the neighborhood\cite{lundberg2017unified} whereas others, including LIME, choose to apply a similarity kernel to determine the weight of each neighbor instance\cite{ribeiro2016should,shrikumar2017learning}.

% Once the local model is built, it is reported as per Figure~\ref{fig:lime}. Attributes ranges
% are sorted according to their feature importance;
% i.e. how much they can change the class label of a test instance $T$.

\subsection{Alternatives to LIME}
As mentioned above,
Mueller et.al.~\cite{mueller2019explanation} 
report that the literature
on explanation is truly vast.
Consequently, there are many alternatives
to LIME including the abductive
framework of Menzies et al.~\cite{menzies02}
or ANCHORS~\cite{ribeiro2018anchors} (which is another
explanation algorithm generated
by the same team that created LIME).
Given that explanation is such a large
field, it is  appropriate to ask
why this paper commits
to the LIME view of explanation
and not some other approach.

Firstly, LIME is operational
whereas much of the (say)
philosophical literature on explanation
is insightful, but not executable.

Secondly, LIME handles an important
detail that other approaches ignore.
As mentioned above by Mueller et.al.,
some discussions on  explanation 
ignore how to formulate problems 
and how to use the explanations to
resolve problems. LIME, on the other hand,
formulates the problem as a data mining task
where ``explanation'' is operationalized
as a regression problem learning gradients
around a point in instance space.
LIME also
offers the following resolution operator:
find attribute ranges that change the
class of an instance into something 
more desirable.

Thirdly, LIME scales to large problems. Much recent work has results in
methods to scale data mining to very
large data sets. Since LIME is based
on data mining, then LIME can use
those scalability results in order
to generate explanations for very
large problems.

Fourthly, and this is more of a low-level
systems reason, alternatives to LIME
such as ANCHORS assume discrete classes.
Our data has continuous classes which could be binarized into two discrete classes-- but only at the cost of losing the information about local gradients. Hence, at least for now, we explore LIME (and will explore
ANCHORS in future work).

Lastly, LIME is a widely-cited algorithm. At the time of this writing, LIME has received over 2,600 citations since it was published in 2016. Hence, methods used to improve LIME could also be useful for a wide range 
of other research tasks.
This paper proposes precedence plausibility as a way to improve LIME.

\subsection{Precedence-based Plausibility}\label{gpr}
%Beyond the simple way to generate recommendations, we introduce  \emph{precedence} analysis into the design of the recommendation generator.
%as we believe that scientific recommendations are those who incorporate with the domain knowledge within that field.  
A workshop on "Actionable Analytics" at ASE'15\cite{hihn2015data} reported complaints from business users about the analytic models such as "rather than apply a black-box data mining algorithm, they preferred an approach with a seemingly intuitive appeal". Since software engineers are the target audience of explanations in SE, it is crucial to ensure the explanations are valued by them.  Chen et al. say the term "actionable" can be defined as a combination of "comprehensible" and "operational"\cite{chen2018applications}. But how to assess "operational"? 

In this paper we make the following assumption about ``operational'': a proposed change to the code is plausible  if it has occurred before. That is, in this work, we claim a plan is the most operational when it has the most precedence in the history log of the project.
% This means that a software analytic tool should not only be understood by developers but also be proved achievable in real world practice. In this paper, we argue that a wise way to examine "actionable" is to look into the future, which is the upcoming release of a project. Then we claim that "actionable" explanations are those endorsed by precedence.
Using this assumption, we can generate operational analytics by:
% We say that a recommendation of new actions (to reduce, say, defects) should have some precedence in the historical records of the project. Specifically:
\begin{itemize}
\item Looking at two releases of a project
and report the attributes that have changed between them; 
\item Next, when generating explanations, we only used those attributes that have the most changes.
\end{itemize}
After conducting a survey on 92 controlled experiments published in 12 major software engineering journals, Kampenes et al.~\cite{Kampenes2007} argues that in SE,  size change can be measured via Hedge's $g$ value\cite{rosenthal1994parametric}:
\begin{equation}\label{g}
    g =(M_1 - M_2)/(S_{pooled} )
\end{equation}
Here,  $M_1$ and $M_2$ are the means of an attribute in two consecutive releases  and $S_{pooled}$ comes from
\ref{spool}. This expression is the pooled and weighted standard deviation ($n$ and $s$ denote the sample size and the standard deviation respectively).
\begin{equation}\label{spool}
    S_{pooled} = \sqrt{((n_1-1)s_1^{2}+(n_2-1)s_2^{2})/(n_1+n_2-2)}
\end{equation}

\section{Measuring Effectiveness:  the $K$-test} \label{K-test}

This paper claims that recommendations based on TimeLIME (that focus on attributes with a history of most change) outperform recommendations generated from classical LIME.
To defend that claim, we need some way to assess different planning systems.

Krishna's $K$-test\cite{krishna2017learning}  uses historical data from multiple software releases to compare the effectiveness
of  different plans $P_1,P_2,....$. The test is a kind of simulation study that assumes   developers were told about a plan at some prior time. After that, the test checks what happens for  code
that was changed in accordance (or in defiance) of that plan.

Since the test is a historical study,
it requires consecutive releases $x,y,z$ of some software system.
These releases are required to contain named regions of code $C_1,C_2,$etc  that can be found in releases
$x,y,z$. For example,   $C_i$ could be an object-oriented  class or a function or a file that is found in all releases.
The $K$-test then assumes that there exists a quality measure $Q$ that reports the value of the regions of named code in different releases. In this study, we will use NDPV (\emph{Number of Defects in Previous Version}) as the quality measure, which is described later in \S\ref{Performance Criteria}. 
Some method is then applied that uses $Q$ to reflect on the releases $x,y$
in order to infer a plan $P_i$ for improving release $z$\footnote{Note the connection here to temporal validation in machine learning~\cite{Witten:2011}. In the $K$-test, no knowledge of the final release $z$ is used to generate the plans.}.

Given the above, the $K$-test collects four quantities:
\begin{enumerate}
\item 
$G_{x,y}$: the list of Hedge's $g$ scores for each feature in release $x,y$ 
\item 
$\Delta_{y,z}$: the delta  
between code $C_i$ in releases $y,z$.
\item
$J_{y,z}= \Delta_{y,z} \cap P_i$: the   overlap between the proposed plan and the code changes;
\item
$Q_z-Q_y$: i.e. the change in the quality of the named
code regions between release $y,z$.
\end{enumerate}
The $K$-test assumes that ``good''  plans have the following property:
\[(Q_z-Q_y) \propto |J_{y,z}| \]
That is, increasing the size of the overlap between the proposed plan and the observed changes is associated with an increase in the quality of release $z$.
That is to say, the $K$-test defines \emph{better} plans as follows:
\begin{quote}{\bf DEFINITION:} {\em
 Plan $P_i$  is ``better''
 that     plan $P_j$   if,
  in release $z$, $P_i$ is
  associated with most
  quality improvements.}
 \end{quote}

\noindent 
For our purposes, the $K$-test procedure in this paper consists of four steps:
\begin{itemize}
    \item Train some black-box classifier on version $x$.
    \item Use the classifier and training data to build the explainer in LIME.
    
    \item  Use the classifier and explainer to generate plans with the aim of fixing bugs reported in version $y$. Note that, in this step,
    TimeLIME will combine the explanations from the explainer and the historical data analysis to generate plans.
    
    \item On the same set of files that are reported buggy in version $y$, we measure the overlap score of each plan and the changes in the version $z$ using the Jaccard similarity function. Meanwhile, we also record the change in the number of bugs between the version $y$ and version $z$.
\end{itemize}

For each instance, we compare the extent of overlap between the recommended plan $P_i$ generated by the planner   and the actual developer action in the next release as $\Delta_{y,z}$ using the Jaccard similarity coefficient. 
\begin{equation}\label{similar}
    J_{y,z}(P_i,\Delta_{y,z}) = (P_i \cap \Delta_{y,z})/(P_i \cup \Delta_{y,z})
\end{equation}
Then we convert the coefficient into percentage as our overlap score. As an example shown in Figure \ref{tab:overlap}, the overlap score is \[3/4 \times 100\% = 75\%\]
Formally speaking, the $K$-test is {\em not}
a  deterministic  statement that some plan will necessarily
improve quality is some future release of a project.  Such deterministic causality is a precisely defined concept with the property that a single counterexample can refute the causal claim~\cite{AAAI_1990}. The $K$-test does \underline{not} make such statements.

Instead, the $K$-test is a statement of historical  observation. Plans that are ``better'' (as defined above)
are those which, in the historical log, have been
associated with increased values on some quality measure. Hence, they have some likelihood (but no certainty) that they will do so for future projects.

\begin{table}
\centering
\refstepcounter{table}
\caption{A toy example of how to compute the overlap score using Jaccard similarity function in Eq. (\ref{similar}). Plans P that match the developer actions are marked gray.}
\label{tab:overlap}
\begin{tabular}{|r|c|c|c|c|} 
\hline
 & AMC      & LOC        & LCOM        & CBO           \\ 
\hline
Current release y     & 0.2                                              & 0.1        & 0.9                                              & 0.5                                               \\ 
\hline
$P_i$ from release z       & {\cellcolor[rgb]{0.753,0.753,0.753}}{[}0.1, 0.3] & {[}0, 0.1] & {\cellcolor[rgb]{0.753,0.753,0.753}}{[}0.2, 0.5] & {\cellcolor[rgb]{0.753,0.753,0.753}}{[}0.7, 0.9]  \\ 
\hline
%$\Delta_{y,z}$
Next release z & {\cellcolor[rgb]{0.753,0.753,0.753}}0.2          & 0.3        & {\cellcolor[rgb]{0.753,0.753,0.753}}0.3          & {\cellcolor[rgb]{0.753,0.753,0.753}}0.7           \\
\hline
Match? & y & n & y & y\\\hline
\end{tabular}
\end{table}
% \begin{figure}[t!]
% \tiny
%     \includegraphics[width=1.0\linewidth,height=1.5cm]{toy.png}
%     \caption{A toy example of how to compute the overlap score using Jaccard similarity function in Eq. (\ref{similar}). Plans P that match the developer actions D are marked gray.}
%     \label{fig:overlap}
% \end{figure}

\section{Experimental Methods}\label{experiment}
The experiment reports the performance of the classical LIME and TimeLIME by comparing the quality of plans recommended by each method.

Firstly, we use an over-sampling tool called SMOTE\cite{chawla2002smote} to transform the imbalanced datasets in which defective instances may only take a small ratio of the population. This was needed since, in many of the prior papers that explored our data,  researchers warn that small target classes made it harder to build predictors~\cite{amrit18}.

Secondly, as discussed above, we  train the predictor $P$ and explainer $E$ on data of version $x$. Then in version $y$ we use the explainer to generate explanations {\em only} on those data that are reported as buggy. We also use the predictor $P$ to determine whether we should provide recommendation plans to the instance. 

Then we measure the overlap score of our recommended plan and the actual change on the same file in version $z$. To do this, only select instances that are defective and whose file name has appeared in all releases of data to be instances in need of recommendations. 

The above steps are used for classical LIME as well as the 
TimeLIME planner proposed by this paper.
In the classical LIME planner, we use the simple strategy which is to change as many features as it can in order to reduce the number of bugs. 
On the other hand, for TimeLIME, we first input historical data from the older release to compute the variance of each feature. Then we selected the top-$M$ features with the largest variance as \emph{precedented} features, meaning any recommendation on other features will be rebutted. After getting recommended plans from both planners, we assess the performance of two planners using the overlap score as described in \S\ref{Performance Criteria}. 

Note that the parameter $M$ can be user-specified and the features may vary with respect to different projects and the releases used as historical data. Here we set the default value of $M$ to be 5, which means only $25\%$ of all twenty features can be mutated. Our results from experiments suggest that $M=5$ is a useful default setting. Future work shall explore and compare other values of $M$.

\subsection{Data}
To empirically evaluate classical LIME vs TimeLIME, we use the standard datasets and measures widely used in defect prediction. In this paper, we selected 8 datasets from the publicly available SEACRAFT project\cite{jureczko2010towards} collected by Jureczko et al. for open-source JAVA systems (\url{http://tiny.cc/defects}). These datasets keep the logs of past defects as shown in Table \ref{tab:dataset} and summarize software components using the CK code metrics as shown in Table \ref{ck}. Note that all the metrics are numerical and can be automatically collected for different systems\cite{nagappan2005static}. The definition and nature of each attribute in the metrics is elaborated by prior researchers Jureczko and Madeyski \cite{jureczko2011significance,madeyski2015process}. Another reason this paper selects these 8 datasets is that they all contain at least 3 consecutive releases, which is required by the evaluation measure described in \S\ref{K-test}.

\begin{table}
\centering
\caption{Defect datasets used in the experiment. The last release of 3 release versions in each project is the validation release in $K$-test.}
\label{tab:dataset}
\begin{tabular}{|l|l|l|l|} 
\hline
Dataset  & Release version & No. of files & Bugs(\%)       \\ 
\hline
Jedit    & 4.0, 4.1, 4.2   & 985          & 233 (23.65)   \\
Camel    & 1.2, 1.4, 1.6   & 2445         & 506 (20.70)   \\
Xalan    & 2.5, 2.6, 2.7   & 2597         & 1209 (46.55)  \\
Ant      & 1.5, 1.6, 1.7   & 1389         & 216 (15.55)   \\
Lucene   & 2.0, 2.2, 2.4   & 782          & 379 (48.47)   \\
Velocity & 1.4, 1.5, 1.6   & 639          & 431 (67.45)   \\
Poi      & 1.5, 2.5, 3.0   & 1064         & 637 (59.87)   \\
Synapse  & 1.0, 1.1, 1.2   & 635          & 136 (21.42)   \\
\hline
\end{tabular}
\end{table}

\subsection{ Learner}
Since the goal of this paper is to examine the performance of the explanation tool rather than the predictive model, this paper takes one classifier and applies multiple explanation algorithms.

Our choice of classifier is guided by the  Ghotra et al. \cite{ghotra2015revisiting} study that explored  30 classification techniques for   defect prediction. They found
that all the classifiers they explored fell into four groups
and that  Random Forest classifiers (RFC) were to be found in their top-ranked group.

A RFC is an ensemble learner that fits a number of decision tree classifiers on different sub-samples of the dataset and generates predictions via average voting from all the classifiers\cite{ho1995random}. It is impossible to visualize a fitted RFC as a finite set of rules and conditions due to the voting process. Therefore, RFC is considered a non-interpretable model. Hence, it is a suitable choice for this study.

\subsection{Explainer and Planner} \label{planner}
Using LIME to generate explanations for each prediction made by the learner model, we transform the explanations into recommendations that are expected to shift the prediction probability from positive (buggy) to negative. We use the default parameter setting of LIME, which is 5000 samples around the neighborhood, and the entropy-based discretizer. The explanation object return by a LIME explainer is a tuple in which each element contains the feature name and the corresponding feature importance. It also provides a discretized interval indicating the range of values during which the feature will maintain the same effect to the prediction result. As described in Algorithm \ref{simpleflip}, the simple planner based on the classical LIME will recommend changes on all features that contribute to the defective prediction. The plan on each recommended feature is in the form of an interval, generated by flipping the discretized interval relative to the midpoint of the feature value range $[0,1]$.  

\begin{algorithm}[h!]
\caption{ClassicalPlanner\label{simpleflip}}
% \footnotesize
\KwData{explanation $e$ // the weighted ranges from   Figure \ref{inside}}
\KwResult{A tuple consisting of intervals of values $v$}
\Begin{
$w, v$ $\gets$ $e$ // split weights $w$ and value intervals $v$ from $e$\\
$i \gets0$\\
\While{$i \leq$ sizeof $(w)$ }{
  \eIf{$w[i] \geq$ 0}{
   $v[i] \gets flip\_around\_mid (v[i])$\\
   }
   {
   pass  // do not propose a change on this feature
  }
  $i \gets i+1$
  
}
return $v$
}
\end{algorithm}

%On the other hand, as demonstrated in Algorithm \ref{refineflip}, the TimeLIME planner chooses to mutate features that are indicated operational according to historical analysis.

\begin{algorithm}[h!]
\caption{TimePlanner\label{refineflip}}
% \footnotesize
\KwData{explanation $e$ from Figure \ref{inside}, precedence parameter $M$, previous release $x$, current release $y$}
\KwResult{A tuple consisting of intervals of values $v$}
\Begin{
$w, v \gets e$ // split weights $w$ and value intervals $v$ from $e$\\
$M \gets$ 5 // the default parameter $M$ is 5 meaning at most 5 features can be changed in the resulting plan\\
$g \gets$ hedge($x,y$) // defined in  \S\ref{gpr} \\
precedented $\gets sorted (g) [0:M]$\\
$i \gets0$\\
\While{$i \leq$ sizeof $(w)$ }{
  \eIf{$w[i] \geq$ 0 and $i \in$ precedented}{
   $v[i] \gets flip\_around\_mid (v[i])$\\
   }
   {
   pass // do not propose a change on this feature
  }
  $i \gets i+1$
}
return $v$
}
\end{algorithm}

Apart from the two planners based on LIME, we also use a planner named RandomWalk as a ``straw-man'' baseline algorithm.  This planner, as shown in Algorithm \ref{randomwalk}, assigns random recommendations to each variable stochastically. In our experiment setting, we set the probability to $0.5$ meaning that all features have $50\%$ chance to be recommended a change.  

\begin{algorithm}[h!]
\caption{RandomWalk\label{randomwalk}}
% \footnotesize
\KwData{ standardized code instance to be explained $c$}
\KwResult{A tuple consisting of intervals of values $v$}
 \Begin{
$(a,b) \gets$ $sorted$(rand(1),rand(1)) // generate 2 random float to form an interval within the range [0, 1].\\
$i\gets0$\\
\While{$i \leq$ sizeof $(c)$ }{
  $p \gets$ rand(1) // generate another random float to determine whether a feature needs to be changed or not\\
  \eIf{$p \geq$ 0.5}{
   $v[i] \gets (a,b)$ // apply the random interval.\\ 
   }
   {
   pass // do not propose a change on this feature
  }
  $i \gets i+1$
}
return $v$
}
\end{algorithm}

\subsection{Performance Criteria} \label{Performance Criteria}
The two performance criteria in this experiment, as described in the \S\ref{K-test}, are the overlap score of individual plans and the number of bugs reduced/added in the next release of the project. The function used for computing the overlap score is the Jaccard similarity function in Eq. \ref{similar}, and the other criterion is measured by the metric NDPV (\emph{Number of Defects in Previous Version}), which returns the number of bugs fixed (or added) in a given file during the development of the previous release. The nature of NDPV and similar metrics have been evaluated by plentiful researchers\cite{jureczko2010using,couto2014predicting,shihab2010understanding,khoshgoftaar1998using}. 

% and we believe this study is the case where NDPV becomes most appropriate in empirically evaluating performance of explanations. 

% Note that since the validation data for each planner are the same, the sum of NDPV in all file records is certainly the same as well. Therefore, our performance assessment will value each algorithm by the distribution of NDPV according to the overlap scores. For example, when a bug gets eliminated in a file, the plan of 80 percent overlap score is better than a plan of 20 percent overlap score since the later one displays little similarity to the actual changes undertaken by developers.

\begin{figure}[!t]
    \centering
    \includegraphics[width=1.0\linewidth,height=6cm]{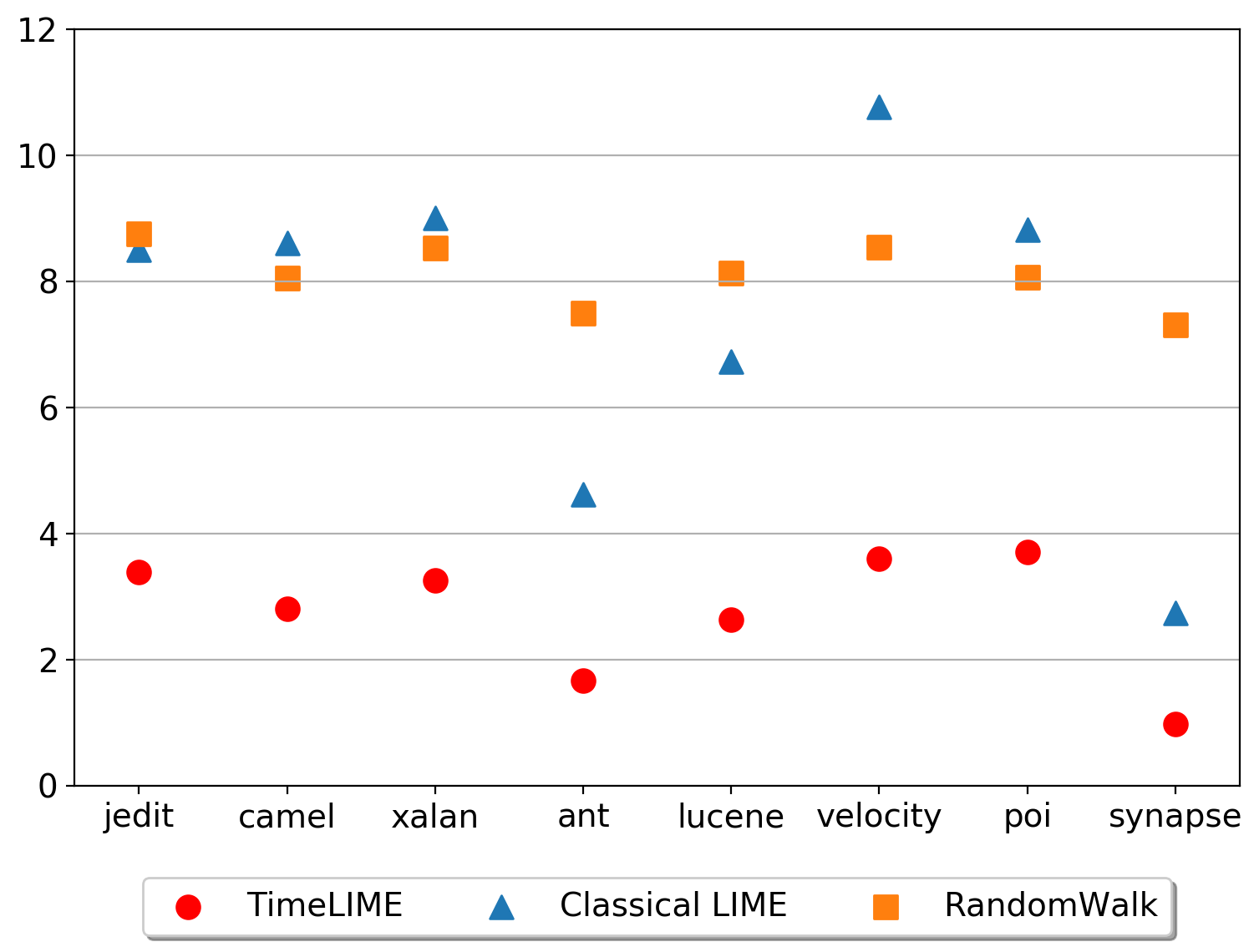}
    \caption{{\bf RQ1 results:} Mean size of plans across all instances in release $z$.   Y-axis= number of features changed by recommended plans. Smaller y-values indicate smaller plans}
    \label{fig:size}
\end{figure}

\begin{figure*}[t!]
    \centering
    \includegraphics[width=1.0\textwidth,height=6.5cm]{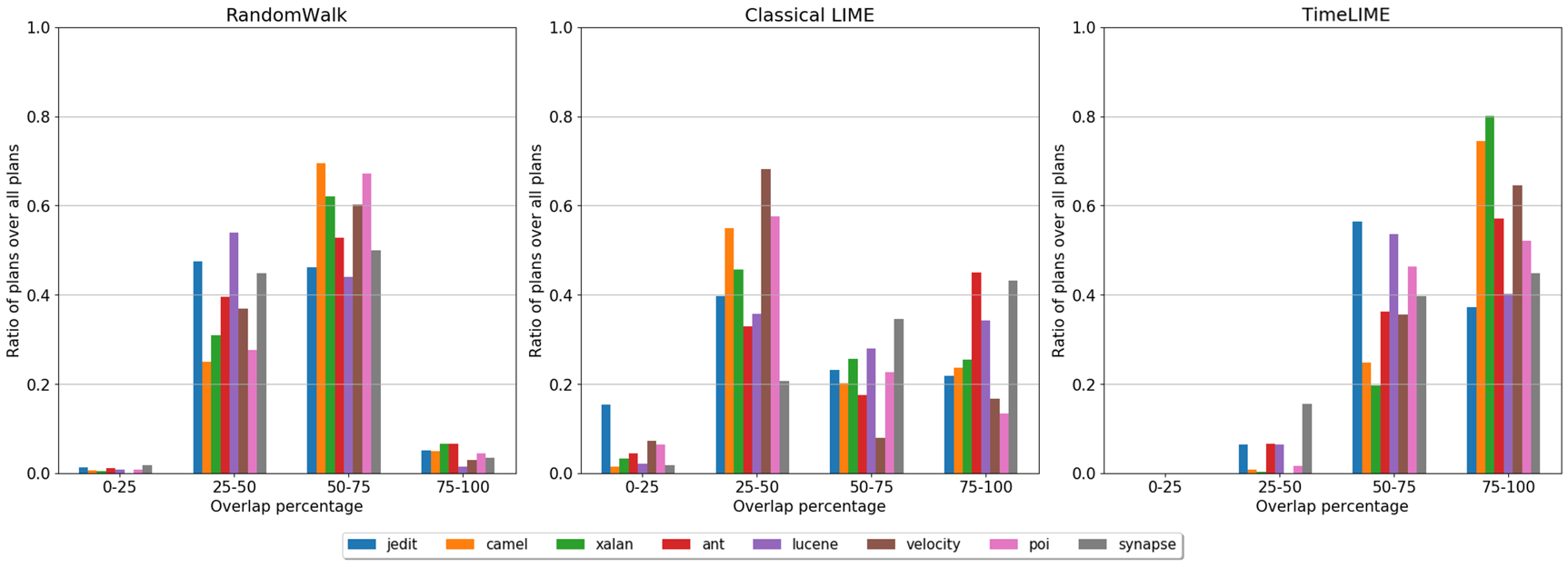}
    \caption{{\bf  RQ2 results:} Distribution of overlap scores of the RandomWalk planner, classical LIME planner, and TimePlanner respectively.   Within each of these three plots,
    results that fall to the right-hand-side  are {\em better} since they
    better correspond to actual developer actions.  Note
    that, by this measure, TimeLIME better reflects
    actual developer changes.
    In this figure, the x-axis is the overlap scores computed by Eq. \ref{similar} and discretized into quantiles. This figure is summarized in
    Figure~\ref{fig:scatter2}.}
    \label{fig:rq2}
\end{figure*}

\section{Results}\label{result}
\textbf{RQ1: Are all explanations precedented?}

Before doing anything else, we need to assess if there are any differences between the explanations generated by TimeLIME and those of classic LIME. This is important to check since if both algorithms are producing the same recommendations, then there is little  point to this paper.

Figure \ref{fig:size} reports the 
mean size of plans across all instances in release $z$. 
In terms of the size of the proposed changes, TimeLIME generates much smaller recommendation plans compared to the classical LIME and random planner. Note that since TimeLIME in the experiment restricts recommendations to the top 5 features with highest Hedge's $g$ scores, the size of an TimeLIME plan will never be more than 5. However, as shown in the figure, the average size of TimeLIME plans is always smaller than 5. This implies that the original explanation sets, returned by the classical LIME, do contain unprecedented explanations which then get rejected by the TimePlanner.   
% \newpage
Hence we say that 
\begin{blockquote}
% [width=\linewidth,colbacktitle=gray,title={Result 1}]   
\textbf{Answer 1}: Widely-used explanation algorithms (classic LIME) do not restrict themselves to explanations with precedence in the historical record of a project.
\end{blockquote}

Note that we view this result as a potential flaw in classic-LIME since, as shown below, better explanations arise from using just the precedented attributes.

 \begin{figure}[]
    \centering
    \includegraphics[width=1.0\linewidth,height=6.5cm]{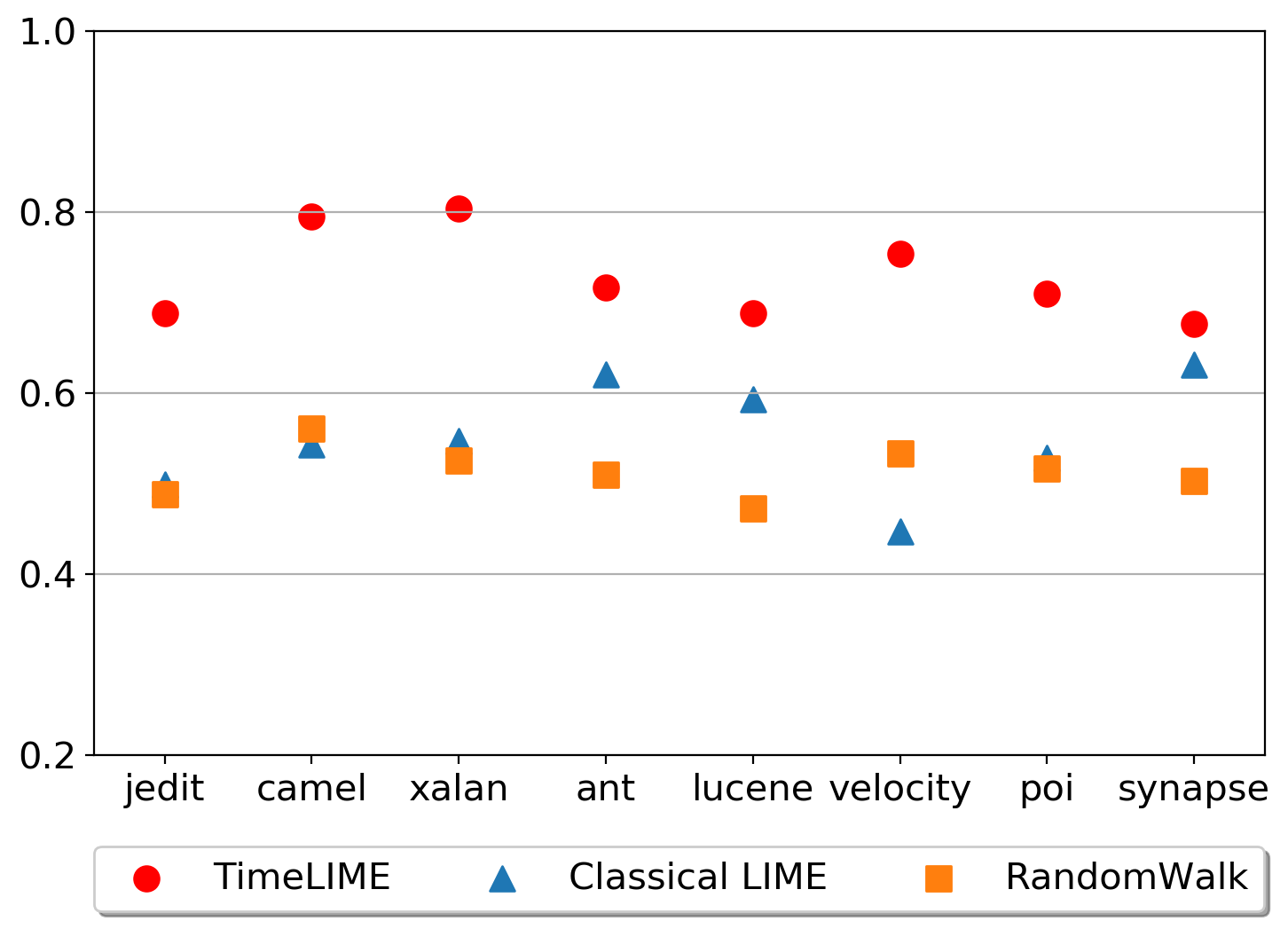}
    \caption{{\bf More RQ2 results:} The mean overlap scores of all recommendation plans made by 3 planners seen in Figure~\ref{fig:rq2}.
   }
    \label{fig:scatter2}
\end{figure}

% According to {\bf DEFINITION~1} from
% \S\ref{K-test},
% the TimeLIME planner
% is ``better'' than the
% Classical-Line planner if 
%   in the final release,
%   the former is
%   associated with most
%   quality improvements.
\textbf{RQ2: Do developers prefer precedented explanations?}

An explanation/recommendation can be proven useful if there is evidence indicating developers could actually apply those kinds of
changes.
Figure \ref{fig:rq2} comments on how often developers are willing to perform the plans suggested by different planners. This figure
was generated using the $K$-test
procedure described above.
The x-axis of that figure shows the 
$J_{y,z}$ overlap measure from Eq. \ref{similar} in \S\ref{K-test}. 
% Within any of the three plots of that figure, results on the right come from changes that most corresponded to some proposed plan. 
The y-axis of that figure shows the portion of plans falling into each overlap score quantile among all plans generated by planners. 
%It represents the discretized frequency of overlap scores. 

In that visual representation, the planners whose actions most correspond to known developer actions have higher values on the {\em right-hand-side} of each plot. Such values illustrate examples where changes proposed by the planner most correspond  to changes made by the developers. In general, we observe the tendency that the TimePlanner generates plans that are much more favored by developers than the plans from  the other 2 planners. 

To facilitate the  comparison over the  planners, Figure~\ref{fig:scatter2} lists three sets of average overlap scores for Random, classical LIME, and Refined-LIME:
\begin{itemize}
\item The classical LIME recommendations do not correspond well with known developer actions (since the expected overlap score within a project is around 0.5, sometimes even lower).
\item
The plans provided by the classical LIME have no significant difference ($p < .05$) from  RandomWalk.
\item
Of the 3 planners studied here, TimeLIME's plans most reflect the actions of developers.
\end{itemize}
% the average overlap scores from three methods on each project. On all projects, the TimeLIME has achieved a higher overlap score than the classical LIME. It can be then indicated that our refined method can generally generate more actionable recommendations based on the same explanations originally provided by LIME. By comparing the classical LIME and the TimeLIME,we find out that in the TimeLIME, plans that are most similar (75\% to 100\% similarity) to real changes tend to reduce most bugs. On the contrary, in most cases, the classical LIME is unlikely to provide recommendations that are similar to the real changes. 
% As described in Figure \ref{fig:total}, in the majority of projects more bugs are reduced than introduced. Since the total number aggregated within each project will not change whereas the distribution will, we believe that the weighted scores, which is computed as the final number of bugs reduced weighted by the similarity scores of each corresponding plans, would generate a more scientific and convincing illustration on the performance assessment. According to Figure \ref{fig:scatter}, we conclude that TimeLIME not only provides more actionable explanations, but also achieves a greater net gain in terms of a higher weighted score. 
\newpage
Hence we say:
 \begin{blockquote}
% [width=\linewidth,colbacktitle=gray,title={Result 1}]   
\noindent
\textbf{Answer 2}: Of all the planners studies here, developers are less likely to perform the plans proposed by classical LIME than TimeLIME. That is to say, the precedented explanations are more favored by developers. 
\end{blockquote}

%  \begin{figure}[h!]
%     \centering
%     \includegraphics[width=1.0\linewidth,height=6cm]{scatter2.png}
%     \caption{{\bf More RQ2 results:} The average overlap scores of 3 planners show that TimeLIME generally provide plans that better match the actual developer actions in the next release. And the overlap scores of plans provided by the classical LIME have no significant difference from the random planner if one compare the scores horizontally across 8 projects.}
%     \label{fig:scatter2}
% \end{figure}

\begin{figure*}[]
    \centering
    \tiny
    \includegraphics[width=1.0\textwidth,height=6.5cm]{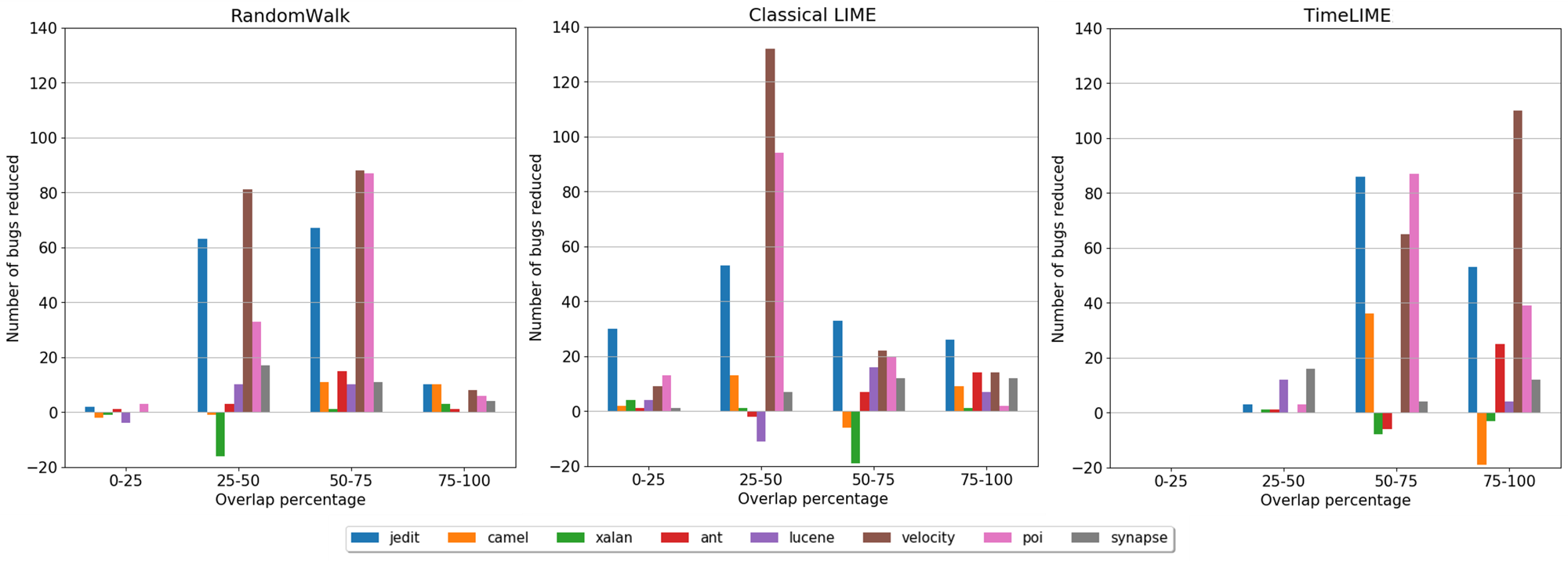}
    \caption{{\bf RQ3 results:} Total number of bugs reduced by the RandomWalk planner, classical LIME planner, and the TimePlanner respectively.
    Within each of these three plots,
    results that fall to the right-hand-side  are {\em better} since they
    they most correspond to most defect improvement
    in the subsequent release.   By this measure, TimeLIME is the {\em best} since
    its plans are associated with most defect reduction.
    In this chart,
    a positive number implies a reduction of bugs in the validation dataset. A negative number means an increase in the total number of bugs. The sum of 4 bars should be the same within each project despite different planners since the same validation datasets are used. This chart is summarised in Figure~\ref{fig:scatter}. }
    \label{fig:total}
\end{figure*}

\begin{figure}[]
    \centering
    \includegraphics[width=1.0\linewidth,height=6cm]{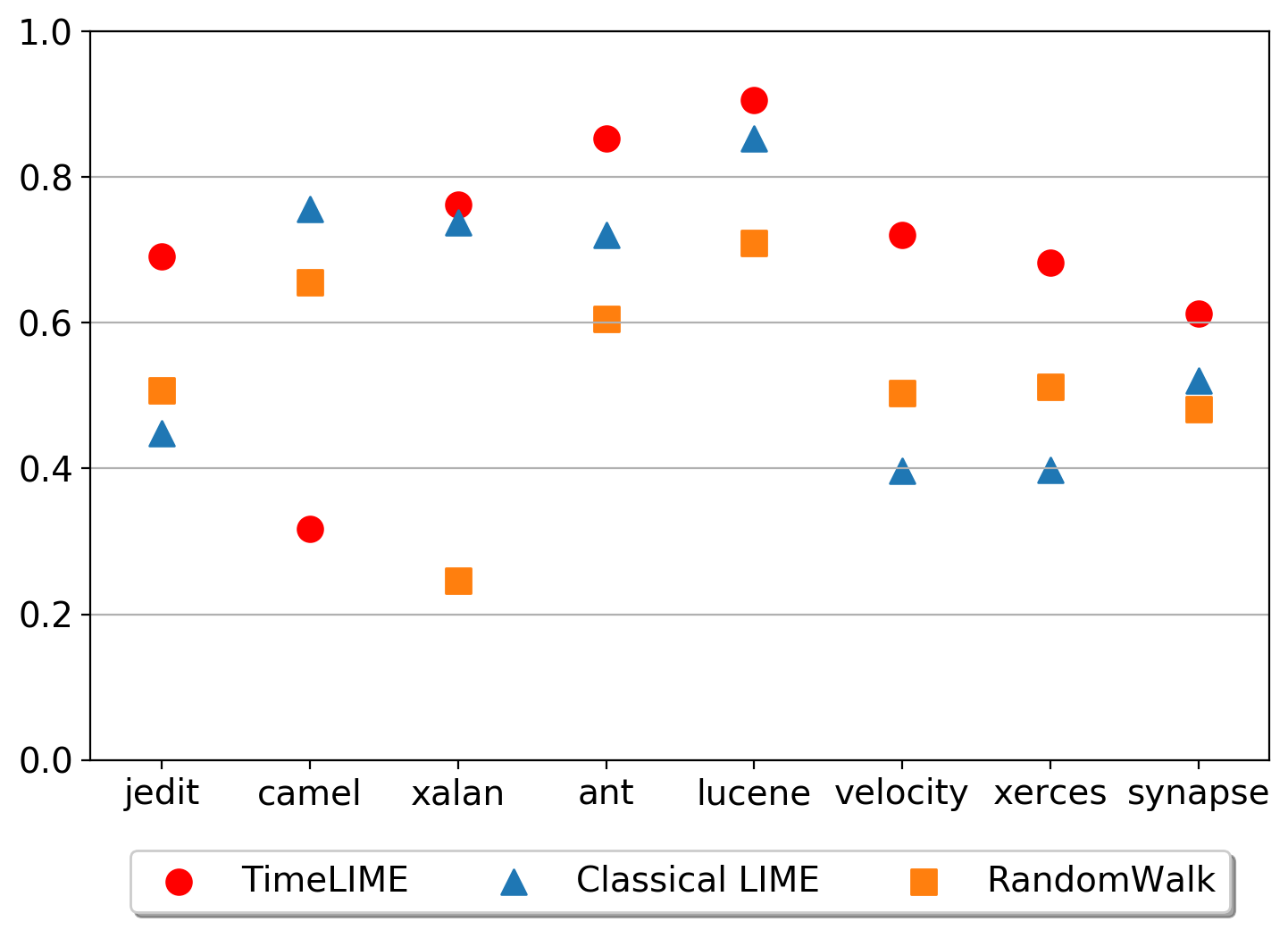}
    \caption{{\bf RQ3 results:} Summary
    of Figure~\ref{fig:total}. The expected values of weighted scores $S_{scaled}$ of 3 planners in each projects as computed by Eq.~\ref{scaled}.}
    \label{fig:scatter}
    \vspace{-.5em}

\end{figure}
% \newpage
\textbf{RQ3: Are precedented explanations better at defect reduction?}

As discussed earlier, better explanations in SE are believed to be explanations that are (a) easier to apply while (b) maintaining the effectiveness in reducing bugs. 

The first criterion has already been met.
As seen there, the recommendations made by TimeLIME
are much smaller, hence easier to apply, than the other
methods studied here. Also, as seen above, the recommendations
from Refined-LIME correspond to the known actions of developers.

To measure the second criterion, we chose to use a weighted sum function to compute the net gain of each planner. The weighted sum function in Eq. (\ref{ws}) weights the NDPV by the overlap score of the plan. 

In the experiment, each plan $p_i$ from the all $N$ plans returns an overlap score $s_i$ and a
NDPV number $n_i$ (positive number indicates bugs reduced, negative number indicates bugs added).
Then we weight the NDPV $n_i$ by the planner by $s_i$ to compute the aggregate score $S$. 
\begin{equation}\label{ws}
    S= \sum {s_i * n_i}
\end{equation}
Note that the {\em larger} the  overlap the {\em greater}
the change in the number of defects introduced. 
  Equivalently, a very high overlap score of a plan that ends up with new bugs added in the next release implies strong unreliability of this plan. As a result, the planner should receive more scores deducted by this plan.

Additionally, given that the total number of bugs varies from each project as shown in Figure \ref{fig:total}, a  project with more bugs reduced in the validation dataset will expect the planner to score more than the planner whose validation dataset has fewer bugs reduced so that their performance can be considered proportionally similar. For example, project A has $NDPV = 100$ in release $y$ and another project B has $NDPV = 10$ in its next release y. If one would like to see similar performance of a planner on these 2 projects, the weighted score in project A $S_{A}$  is expected to be 10 times higher than $S_{B}$ since there are potentially more bugs  that can be reduced by a planner in project $A$ than in project $B$ and it won't make any sense if a planner gains the same score in both projects.  
From this perspective, we scale the final score $S$ in Eq. \ref{ws} by the sum of NDPV within the project to get the scaled score $S_{scaled}$.
\begin{equation}\label{scaled}
    S_{scaled} = \frac{\sum_{i}^{N} {s_i * n_i}}{\sum_{i}^{N} {n_i}} 
\end{equation}
% By measuring the scaled score, we only evaluate the performance among different planners, but also compare the performance of the same planner among different projects in order to assess the robustness of each planner based on the fluctuation of the scaled final score.  
The visualized result in Figure \ref{fig:scatter} shows that the TimePlanner obtains highest average $S_{scaled}$ scores in most of the projects (7 out of 8).  

As to the one case that failed (CAMEL), we have investigated various reasons why that might be so. Looking at the distributions of its features, we cannot see anything that distinguishes CAMEL from the other projects.  The most promising possibility is that the staffing profile of CAMEL changed dramatically during the releases studied here, which means that numerous extra bugs arrived due to the inexperience of the new staff. 

Whatever the reason for the CAMEL result, the overall result
is very clear: 
\begin{blockquote}
    \noindent
    \textbf{Answer 3}: TimeLIME's precedented explanations are associated with  greater defect reduction.
    They are also easier for developers to apply.
    \end{blockquote}

\section{Threats to validity}\label{threat}
Due to the complexity of the experiment designed in this case study, there are many factors that can threaten the validity of these results.
\subsection{Learner bias} 
This paper selects RFC as the black-box classifier because prior research has shown that RFC is ranked as one of the top models among all 32 classifiers used in defect prediction. However, the preeminent predictive power of RFC does not ensure that explanations derived from it are preeminent recommendations as well. Other methods from the top rank may be more suitable in the problem of explanation generation while we haven't explored more.

\subsection{Instrument bias}
Explainable AI is experiencing its resurgence and various approaches are proposed to generate explanations. Although LIME is one of the widely cited and well-known tools, it is possible other tools are more suitable in solving SE problems, which can make solutions from LIME sub-optimal. Hence, to verify if adding in SE knowledge can always improve AI tools, we need to make a comprehensive exploration that includes more explanation generation methods.
% \subsection{Hyper-parameter Tuning}
% Past researches have shown how hyper-parameter optimization can boost the performance of a classifier used in defect prediction. Since in this paper we concentrate on the modification of the explainer instead of the learner, we used a simple grid search to find the optimal parameter setting. It can be possible that the current setting is sub-optimal and by using the actually optimal settings we might receive different experiment result.

\subsection{Evaluation bias}
Experimentation in this paper uses performance measures as defined above. Other similarity score functions are also widely used in research. A comprehensive analysis using these measures
can be further performed using our replication package. Additionally, other measures can easily be added to extend this replication package.

\subsection{Sampling bias}
This paper uses historical data analysis to restrain recommendation generation in which 3 releases are collected per project. However, we still prefer to collect more releases of the project and augment the historical data analysis. Recent research in defect prediction has revealed that among several past releases of the project, there exists one bellwether release that is the most suitable training dataset\cite{krishna2018bellwethers}. Therefore, we have reasons to believe in a similar conjecture that there exists such bellwether release that is most helpful in fitting the learner and explainer. 

\section{Future work}\label{futurework}
For future work, we need to take action to retire the above threats
to validity. 
\subsection{More Learners}
More black-box learners should be used in the experiment to construct a more comprehensive comparison. Although the limited sample amount of defect prediction datasets has ruled out many deep learning models such as Neural Network due to the overhead, there are still many other models, including but not limited to Random Subspace Sampling and Sequential Minimal Optimization, applicable for this experiment.
\subsection{More Explainers}
As described above, LIME is a representative member in the family of local surrogate interpretation models. Other local explanation generation methods that apply tree-structure extraction or association rule mining or so on should also be introduced in the discussion.
\subsection{More Data}
We would like to collect not only more SE projects of defection prediction data but also more releases of a single project. This can facilitate the further exploration on the accountability of our historical data analysis.

\section{Conclusion}\label{conclusion}

When dealing with temporal data (e.g., successive software releases), it is useful to restrict any conclusions to actions that have appeared in the historical record of that project. 
This paper has compared  planners built upon the classical LIME explanations that do/do not respect temporal
precedence. We find that plans that respect
precedence: 
\begin{itemize}
    \item \emph{Are smaller}: In terms of the average size of recommended plans. The TimeLIME generally generates smaller plans than the classical LIME and RandomWalk in every project.
    Smaller plans are preferred to larger plan since the latter can be faster to apply.
    \item \emph{Are preferred by developers}: In terms of the overlap between the proposed plans and the developer actions in the upcoming release, plans proposed by TimeLIME better match what developers actually do.
    \item \emph{Are better}: In terms of the scaled weighted scores $S_{scaled}$ that indicate the overall net gain received per project. TimeLIME gets the highest score among 3 planner in 7 out of 8 projects 
    (while the classical LIME wins in only 1 project).
\end{itemize}

In conclusion, we assert two things.
Firstly, the above results clearly show that precedented explanations lead to better explanations (and better plans based on those explanations). 

Secondly, and more generally,  our community should be more careful about using off-the-shelf AI tools without first adapting them using SE knowledge. We think it is rash and ill-advised just to throw standard AI tools at   SE problems.  Those AI methods can be greatly enhanced via SE knowledge.
As shown here, adding that knowledge is   not  a complex thing to do.   Further, once that knowledge is applied, this can result in dramatically better systems.

\section*{Acknowledgements}
This work was partially funded by 
a research grant from the Laboratory for Analytical Sciences, North Carolina State University.

% blinded for review.

\bibliographystyle{ACM-Reference-Format}
\bibliography{reference}

\appendix

\end{document}